\documentclass[sigconf]{acmart}

\AtBeginDocument{%
  \providecommand\BibTeX{{%
    \normalfont B\kern-0.5em{\scshape i\kern-0.25em b}\kern-0.8em\TeX}}}

\usepackage{caption}
\usepackage{subcaption}
\usepackage{array}

\copyrightyear{2024}
\acmYear{2024}
\setcopyright{rightsretained}
\acmConference[CHI '24]{Proceedings of the CHI Conference on Human Factors in Computing Systems}{May 11--16, 2024}{Honolulu, HI, USA}
\acmBooktitle{Proceedings of the CHI Conference on Human Factors in Computing Systems (CHI '24), May 11--16, 2024, Honolulu, HI, USA}
\acmDOI{10.1145/3613904.3642099}
\acmISBN{979-8-4007-0330-0/24/05}

\newcolumntype{P}[1]{>{\centering\arraybackslash}p{#1}}
    
\graphicspath{{images/}} 
\usepackage{ifthen}

\begin{document}


\title[eKichabi v2]{eKichabi v2: Designing and Scaling a Dual-Platform Agricultural Technology in Rural Tanzania}

\author{Ananditha Raghunath}
\authornote{Both first authors contributed equally to the paper}
\affiliation{%
 \institution{University of Washington}
 \city{Seattle}
 \state{Washington}
 \country{USA}}
 \email{araghu@cs.washington.edu}

\author{Alexander Metzger}
\authornotemark[1]
\affiliation{%
 \institution{University of Washington}
 \city{Seattle}
 \state{Washington}
 \country{USA}}
 \email{metzgera@cs.washington.edu}

\author{Hans Easton}
\authornote{All second authors contributed equally to the paper}
\affiliation{%
 \institution{University of Washington}
 \city{Seattle}
 \state{Washington}
 \country{USA}
} \email{hans00@uw.edu}

\author{XunMei Liu}
\authornotemark[2]
\affiliation{%
 \institution{University of Washington}
 \city{Seattle}
 \state{Washington}
 \country{USA}
} \email{xunmel@uw.edu}

\author{Fanchong Wang}
\authornotemark[2]
\affiliation{%
 \institution{University of Washington}
 \city{Seattle}
 \state{Washington}
 \country{USA}
} \email{fw29@uw.edu}

\author{Yunqi Wang}
\authornotemark[2]
\affiliation{%
 \institution{University of Washington}
 \city{Seattle}
 \state{Washington}
 \country{USA}
} \email{yunqiw681@uw.edu}

\author{Yunwei Zhao}
\authornotemark[2]
\affiliation{%
 \institution{University of Washington}
 \city{Seattle}
 \state{Washington}
 \country{USA}
} \email{yzhao4@uw.edu}

\author{Hosea Mpogole}
\affiliation{
  \institution{Institute of Rural Development Planning}
  \city{Dodoma}
  \state{Dodoma}
  \country{Tanzania}
} \email{hmpogole@irdp.ac.tz}

\author{Richard Anderson}
\affiliation{%
 \institution{University of Washington}
 \city{Seattle}
 \state{Washington}
 \country{USA}}
\email{anderson@cs.washington.edu}

\renewcommand{\shortauthors}{Raghunath \& Metzger et al.}

\begin{abstract}
Although farmers in Sub-Saharan Africa are accessing feature phones and smartphones at historically high rates, they face challenges finding a robust network of agricultural contacts. With collaborators, we conduct a quantitative survey of 1014 agricultural households in Kagera, Tanzania to characterize technology access, use, and comfort levels in the region. Recognizing the paucity of research on dual-platform technologies that cater to both feature phone and smartphone users, we develop and deploy eKichabi v2, a searchable directory of 9833 agriculture-related enterprises accessible via a USSD application and an Android application. To bridge the gap in affordances between the two applications, we conduct a mixed methods pilot leveraging mobile money agents as intermediators for our USSD application's users. Through our investigations, we identify the advantages, obstacles, and critical considerations in the design, implementation, and scalability of agricultural information systems tailored to both feature phone \textit{and} smartphone users in Sub-Saharan Africa.

\end{abstract}


\maketitle
\section{Introduction} 

Establishing robust, sustainable, and equitable food systems is imperative to nourish an anticipated population of 10 billion individuals by 2050 \cite{worldbank}. In Tanzania, 80\% of adults are smallholder farmers who face a myriad of challenges within the evolving global context \cite{evans}. One such challenge is their vulnerability to high information friction. For example, networks of rural farmers and surrounding agricultural businesses remain small due to significant barriers associated with acquiring new agricultural contacts. Over the course of our fieldwork in Kagera, Tanzania, we were told several stories that characterize the negative impacts of small network sizes. For example, a widowed subsistence farmer routinely relied on a transporter to drive through her village and buy her surplus harvest. This was her only way to get cash to use in the formal economy. Lacking connections to alternative transporters and being located far from the market, she had no avenues to sell her goods if the transporter chose not to come by. When he did come, she didn't have any leverage to negotiate with him and often received unfairly low prices for her goods. Although she had owned a feature phone for many years, her only contacts were of her children that lived in the capital city. Her story, and many similar ones, emphasize the need to create and maintain agricultural directories that farmers can use to find contacts for agriculture-adjacent businesses. 

To this end, in 2015, Blumenstock et al. created, printed, and disseminated a paper telephone directory of nearby agricultural businesses in central Tanzania titled \textbf{Ki}tabu \textbf{cha} \textbf{Bi}ashara (\textit{Kichabi}) \cite{BlumenstockAkerDillon}. They conducted an RCT to measure the causal effects of having access to nearby contacts on farming households (HHs). They found that "enterprises listed in the directory saw increases in customer contact, sales, and employment ... and agricultural HHs that received directories were more likely to rent land and hire labor, had lower rates of crop failure, and sold crops for weakly higher prices.” Although there were significant gains associated with Kichabi, the paper directory was hard to distribute and keep up to date, showing promise for a mobile phone based agricultural information service that could scale and adapt to the changing business landscape. Building upon Blumenstock et. al’s work, Weld et. al. piloted a USSD-based directory \textit{eKichabi} in 2018 housing 500 businesses deployed to 107 users \cite{weld2018ekichabi}. USSD is a popular protocol in SSA, letting users dial a shortcode to initiate a session with their mobile network operator (MNO) without relying on internet connectivity. While the eKichabi pilot  demonstrated the feasibility and potential for USSD technology to support an agricultural directory, challenges with USSD-based information services \textit{at scale} remain unexplored, i.e., usability challenges with a growing number of diverse users (deployment scale) and development challenges with a larger set of firms listed (application scale). 

Since the 2018 eKichabi pilot, smartphone ownership has grown in Sub Saharan Africa (SSA), with 48\% of SSA inhabitants having smartphone access in 2021 \cite{GSMA2021}. In response, governments and developers are increasingly creating smartphone apps to disseminate agricultural information. Numerous affordances of smartphones— such as image capture, global positioning systems, Internet connectivity, and touch screens— show promise to support users with differing levels of comfort navigating digital services. Currently, designers build either smartphone apps that leverage affordances available to an increasing proportion of users or feature phone apps that forego such affordances to serve the majority of users. The design and implementation of \textit{dual-platform technologies}, i.e. technologies with different versions designed for and accessible to both smartphone and feature phone users, has been overlooked. 

Our research documents the challenges and rewards in creating dual-platform technologies at scale. The research questions we explore address how device ownership and comfort changes in different demographics, how two versions of a digital tool rooted in different modes of technology diverge from one another; how farmers' use and preferences for agricultural information services change on different platforms; and where gaps in affordances between platforms can be bridged through intermediation. Our contributions follow. 

\begin{enumerate}
    \item Through a survey of 1014 HHs in Kagera, Tanzania, we shed light on feature and smartphone ownership, USSD use, and self-perceived comfort levels with USSD technologies. \textit{We uncover a persistently low smartphone ownership in this rural area, a large minority of non-users of USSD, a broad diversity of self-perceived comfort in technology use, and sustained utilization of mobile money agents (wakalas in Swahili) as intermediaries for technology.} 
    \vspace{1em}

    \item We design and develop a dual-platform agricultural directory that works on feature phones (a complete redesign of Weld et al.'s pilot agricultural USSD app) and smartphones (through an offline Android app). We report on usability and trust-building adaptations discovered through user testing, \textit{finding that the inclusion of business owner names greatly increases trust in the directory.} As we scale from 500 to 9833 businesses represented, we describe inherent divergences, limits and capabilities between USSD and offline Android applications, finding \textit{USSD to be increasingly limited in its ability to deliver utility to users with increased application \& deployment scale.}
    \vspace{1em}
    
    \item We deploy the dual-platform technology to 1014 HHs and analyze use patterns across demographics. \textit{We uncover diverse use and preferences among users within each platform. We uncover reluctance to download Android apps, causing both smartphone and feature phone users to rely heavily on USSD. However, we see increased complexity of use enabled by the Android app with lower user effort, leading to greater information gain and retention in Android app users.} 
    \vspace{1em}
    
    \item For those who lack smartphone access, we conduct a small pilot investigation that uses wakalas, whom we train as intermediaries to bridge the gap in usability and affordances between USSD and Android apps. \textit{We discover that wakalas are capable of and interested in intermediating, and community members engage and show greater trust in and use of the USSD app as a result, highlighting important considerations for future iterations of dual-platform interventions.}
\end{enumerate}

\section{Related Work}

We situate our work in related literature on mobile phone based interventions in agriculture, covering a range of technologies and benefits. We also discuss intermediation practices that have been tested to bridge the gap to information services for lower literate and lower comfort mobile phone owners. 

Over the past two decades, Tanzania’s feature phone penetration has increased from 0\% to 101\% \cite{TCRA}, while more than 65\% of the county's labor force has continued working in agriculture \cite{FAOandITU_2022}. A substantial body of literature recognizes that mobile phones hold significant potential to support farmers in SSA \cite{ valtonen2015impact, aker2011dial, aker2010mobile, courtois2015farmer, burrell2015myth, debo2013leveraging, jagun2008impact, overaa2006networks}. The cost of obtaining information through mobile phones is substantially lower compared to other sources \cite{aker2011dial, aker2010mobile, courtois2015farmer}, and the uncertainty associated with travel delays and changes in demand for goods is decreased, thereby avoiding costly stock outs, unproductive trips \cite{aker2010mobile, burrell2015myth, debo2013leveraging, jagun2008impact, overaa2006networks}, and over-reliance on middlemen to set the terms of trade \cite{Dillon2017HowCA, courtois2015farmer}. However, the literature lacks a granular examination of who is able to benefit from these services, i.e. the demographics of device ownership, comfort, and subsequent use of agricultural interventions in rural SSA \cite{aker2010mobile}. We enrich this body of work with a characterization of the technology landscape in Kagera, Tanzania, emphasizing the comfort with and preexisting use of technology as it applies to agricultural activities.

\subsection{Technologies for Agricultural Information}
Research has explored the aptness of varied mobile technologies for disseminating information in SSA —  Short Message Service (SMS), SIM apps, Unstructured Supplementary Service Data (USSD), etc. The SMS interface has been considered for web search and agriculture information services \cite{medhi2011designing, chen2010sms, schusteritsch2005mobile, yi2008deciphering},  but it was found to have limited impact \cite{burrell2015myth, fafchamps2012impact}; the technical challenges, e.g., SMS is error-prone \cite{pervaiz2015user} and subject to delays, costs and a lack of inherent sessions, make it unsuited for stateful interactions \cite{wyche2016don}. Solutions such as SIM apps and SIM overlays require costly custom hardware development and distribution and have therefore been discarded. 

USSD technology has been instrumental in mobile money systems such as mPesa (used by 57 million SSA residents in 2023 \cite{mpesa}) and for providing government services such as MSeva in India \cite{sanka2019electronic, bourreau2015competition}. It offers advantages, such as: a session-based protocol that supports multi-step operations and stateful interaction; familiarity among users through their existing experience with mobile money systems; compact packet size, leading to faster delivery than SMS; and message order guarantees with higher security due to persistent sessions. As the Tanzanian technology landscape changes, and more than 2.6 billion USD has been invested in network infrastructure improvements \cite{FAOandITU_2022}, new device owners stand to benefit from USSD-based information services. Previously, development was limited by carrier control, requiring negotiations with individual carriers to implement a third-party USSD service. However, the number of third-party USSD providers (such as Niafikra) has increased in recent years, with stable gateways available in Kenya, South Africa, Tanzania, and Nigeria leveraged by existing work \cite{weld2018ekichabi, perrier2015ussd, wyche2012dead, silver2018internet, weld2018ekichabi}. The previous eKichabi project demonstrated the feasibility and potential for USSD technology to support a directory with 500 businesses, having \textasciitilde 2,000 sessions and \textasciitilde 10,000 screen views \cite{weld2018ekichabi}. We add to this body of work, testing the boundaries and capacity of USSD technology to serve as the foundation for a much larger application scale (directory of \textasciitilde 10,000 firms) and a significantly increased deployment scale (user population of 1014 users) than has so far been attempted for an information service. Because ICTs in agriculture are not typically conducted on such a large scale, it has been challenging to distinguish variations in use across sub-populations \cite{aker2015economic, aker2010mobile}. Our work analyzes use by the 1014 users stratified by several demographics factors, including self-assessed comfort with technology. 


Smartphone technologies have also been shown to hold great potential for economic interventions given affordances absent in feature phones \cite{silver2018internet, androidOnlineOffline, FAOandITU_2022}. A plethora of Android-based agriculture applications in India showcase the benefits of a GUI, sensors, bigger screens, increased processing, and more \cite{androidAgricultureIndia, androidAgriInfoIndia}, which motivate expanding the limited research on direct use of smartphones for agricultural interventions in more rural developing regions such as Tanzania. One extant app is Ugani Kiganjani, a simple and affordable mobile and web-based digital agricultural extension service available for weather, crop calendars, nutrition, livestock and fishery sub-sectors in Tanzania, created and maintained by the Food and Agriculture Organization (FAO) \cite{UganiKiganjani}. Digital contact systems with a focus on the smartphone platform have been shown to help circumvent traditional gendered communication practices and empower female farmers to access resources even when prevented traditional access to local organizations/networks \cite{Pfavai2023Women}. Recent HCI research shows the smartphone's effectiveness in encouraging women to start businesses \cite{Raghunath2023Grasshoppers, kapinga2017women, RePEc:unu:wpaper:wp-2022-67, Mashenene2014women} with the caveat that it can make them more vulnerable to economic exploitation when not taken into account during the design \cite{Raghunath2023Grasshoppers}. An HCI survey on smartphone usage for African populations \cite{Sakpere2021AgeDifferences} found that when designed to prevent problematic overuse and encourage face-to-face interaction, mobile applications can have a positive impact across a wide age range and varying professions and levels of education. At the same time, the disproportionate cost of smartphone data compared to the low income of significant populations in Tanzania mean that lower income individuals see less benefits of smartphone applications and, from an HCI perspective, care must be taken to provide cheap solutions and understand/design to support the varied use patterns across the range of incomes \cite{mpogole2008poverty}. We enrich this literature by creating a dual platform intervention, making an Android app available that mirrors our USSD app functionality while leveraging smartphones’ affordances. We report on design challenges and the complexities in use enabled in the smartphone-accessing population while building on findings from existing literature, e.g., the importance of offline functionality \cite{androidOnlineOffline}.

Despite the increasing number of platforms that offer agricultural information, the use of these services and access to subsequent extension services are mixed due to socio-technical challenges \cite{magesa2014access, sanga2016mobile, msoffe2019contribution, mwantimwa2019ict, quandt2020mobile, katusiime2021mobile, mushi2022digital, aker2015economic}. In SSA, mobile phone use is hampered in some sub-populations by the cost of acquiring phones, charging the devices, maintaining airtime, repairing phones, and network coverage issues \cite{ahmed2016privacy, wyche2013powering, wyche2015if}, as noted in Kenya \cite{oduor2014technology, wyche2016mobile}, Uganda \cite{martin2011mobile}, and Malawi \cite{steinfield2015mobile}. Another limitation is the farmers' lack of technology knowledge, skills or literacy \cite{wyche2016don, magesa2021smallholder, nyamba2021socio}, and trust in technology\cite{aker2016promise} . Aker et al. \cite{aker2016promise} review various attempts to use agricultural ICTs in Africa and offer sociological and anthropological insight into how these platforms alter the relationship between the provision of information and trust. For rural communities in Tanzania, trust takes different forms \cite{woldmariam2014towards, ali2020two,  raphael2016risks}. Our work documents localized ways to confer trust through design that we learned through iterative user testing.

In an attempt to further bridge some of the aforementioned sociotechnical gaps, \textit{intermediation}, or the use of technology with the help of a trusted, local agent, has shown promise in resource-constrained areas \cite{sambasivan2010human, sambasivan2010intermediated}. Intermediation has been applied to more effectively engage low-literate users \cite{medhi2011designing, lerer2010evaluation, mehra2018prayana, tandon2019designing}; motivate trying new services \cite{ramachandran2010mobile, ramachandran2010research}; facilitate sustained use \cite{katule2016leveraging}; and  enable those with indirect access to technology to benefit from applications \cite{parikh2006intermediation}. 

Some technologies are built specifically with intermediation as the predominant use case, conferring the “primary user” label to the intermediary \cite{parikh2006understanding, mehra2018prayana,kumar2015mobile, wahid2011unlikely, ogbonnaya2019towards}. For example, Rahman et al. \cite{rahman2021dakter} show that small businesses can serve as ICT intermediaries, and that engagement with the small business owners can be extended to engagement with the ICT. Similarly, access to mobile phone-based money services in Tanzania is widespread, with growth attributed to intermediation by mobile money agents \cite{ephraim2016next}. These agents provide essential frontline financial services, including depositing and withdrawing cash, transferring funds, and opening bank accounts, for the rural population \cite{caputo2019pivotal}. Raghunath et al. \cite{Raghunath_2023} uncover that these mobile money agents in Tanzania are losing financial stability when intermediating for extant digital financial services and are interested in creating new channels of income by intermediating on behalf of other ICTs. Based on these findings, we posit that offering intermediation to feature phone app users could serve as an effective means to make up for the lack of affordances of feature phones and circumvent some technology literacy challenges. Our work contributes the first pilot that trains mobile money agents to intermediate for a new ICT, identifying the benefits and challenges as a precursor to a larger scale study. 


\section{Methods}

Our team, composed of researchers from the United States and Tanzania, conducted our mixed-methods study under the umbrella of a large, long-term economic intervention led by academic collaborators in Economics.  The broader project was led by economists with decades of experience implementing agricultural studies Tanzania and was supported by a long term partnership with a Tanzanian research organization. The researchers on this paper include senior authors with substantial experience implementing technology studies in Africa and Asia, as well as undergraduate computer science students with interest in global development projects.  The team brings together researchers with backgrounds from Europe, Africa, South Asia,  East Asia, and North America.

Our joint research undertaking entailed a survey of firms and a survey of HHs across Kagera, followed by development, user testing, and deployment of the dual-platform technology. Finally, we piloted structured intermediation of the USSD technology using mobile money agents. Working with collaborators allowed our reach to extend to a larger scale than otherwise achievable; however, it also limited our contribution to some methodological components (e.g., collaborators had already established power calculations and sampling strategy for households being surveyed in relation to their chosen economics outcomes) or downstream analysis (e.g., collaborators will measure long-term impacts, changes in agricultural productivity, etc., due to the agricultural information service). Below, we provide in-depth descriptions of the methods falling under our purview. We comment briefly on collaborators’ methods where necessary, but refer readers to their forthcoming publication for additional details. All research was approved by our respective IRBs.

\subsection{Firm and Household Surveys}
\label{information_collection}
Author 1 and collaborators moved to Kagera, Tanzania in May 2022 to begin study activities. In June 2022, our collaborators designed and implemented a firm survey in 100 villages in the Kagera region to collect contact information, goods sold, and other relevant information from 9833 agriculture-related businesses across 100 villages in Kagera. This survey formed the basis of our digital directory. The household (HH) survey was conducted in the same 100 villages concurrent with our apps' roll out in November 2022. Our team embedded questions about phone ownership, USSD comfort, intermediated technology use, etc., into the HH survey. Local enumerators visited HHs sampled through random walks within the villages; we believe that this strategy is uncorrelated with the underlying characteristics of the villages. Both firms and HHs were not incentivized to participate in the surveys. 

\subsubsection{Data Analysis}
The survey data was manually cleaned (duplicates dropped, incorrect information investigated and replaced) and subsequently verified by the authors, testers, and collaborating economists and analyzed in Python. Python is widely used for data processing, visualization, and analysis due to its vibrant data science community and consequent plethora of statistics and graphing libraries. Moreover, the authors were familiar with the language and ecosystem from developing the eKichabi server. Descriptive and inferential statistics (in the form of ANOVA significance tests for categorical variables and t-tests for numerical means) were calculated to support inferences from the HH survey. Sampling across Kagera yielded 1014 HHs surveyed across 100 villages. Average participant age was 51.41 years (sd = 15.41 years), and 44\% were female (n=446). On average, male participants (mean age=48.41 years, std=15.24) were younger than female (mean age=55.22 years, std=14.80).

\subsection{Development and Deployment}
\label{dev_and_dep}
We build an entirely new directory architecture with both USSD and Android components, scaling the application data 20x and the deployment user population 10x across 20x the amount of villages in comparison to Weld et al.'s 2018 work \cite{weld2018ekichabi}. Our iterative app development process involved several rounds of informal and formal user interaction and testing with the USSD and Android apps between May and August of 2022. We convenience sampled participants for these user tests with the help of a local guide during the administration of the firm survey. This yielded 56 participants across 4 villages. 

During each user testing session, the enumerator and first author met with groups of 4 participants. They first explained the project and led participants through two example use-cases on the USSD and Android apps: ("Imagine you need to find a transporter for some goods; this is how you'd do it using our digital directory." AND "Imagine you need the phone number of \textit{"Specific Shop"} in your village center; this is how you'd find it."). They asked participants for one alternative use-case and observed their independent use of the tool. Participants were then asked questions about searches and UI preferences that informed specific app design choices. For example, they were told to imagine a village grocery store and asked how they would identify the store (by the business name, owner, location, storefront, etc.). Finally, enumerators answered questions, inquired about ease of use and perceived utility of the tool, and let participants know that the service would be started in November 2022. 

Our collaborators partnered with Niafikra (a third-party USSD provider) in June 2022 to obtain a USSD code from the Tanzanian government. Once the USSD and Android apps were developed, logging and whitelisting mechanisms were appended before the apps were deployed. The logging system collected all USSD server hits and actions such as favoriting, calling, and searching on the Android app. The whitelist was implemented to manage experimental and control arms for the collaborators' study. Deployment began in November 2022, concurrent with the household survey, where users gave consent and were trained to use the USSD/Android app. (A condensed version of the training is in Appendix A.) All households in the digital study were given access to both the USSD and Android app and directed to use whichever they preferred and could access. 

\subsubsection{Data Analysis}
Field notes from the user testing sessions were transcribed and inductively and deductively coded twice by the authors in Atlas.ti to identify modifications that would broadly increase usability. The final codebook contained the following codes: selection-based search, typed search, trust-related inquiries, imagined benefits, potential features for later iterations, and general inquiries. Quotes are incorporated into our results when we describe iterative improvements that were made. Upon deployment, usage logs were analyzed using Python. User logs were cleaned by identifying outliers from box-plots (1.5 inter-quartile range from the first and third quartile), investigating their cause and dropping data that came from sources that did not fit the study parameters, e.g., testers and author usage logs, failed sessions from non-whitelisted phone numbers, and so on.

\subsection{Intermediation Pilot}
Carrier-specific character limits (usually 160 characters), carrier-specific session timeouts (usually 180 seconds), and small/low-res/low-contrast/low-brightness screens of feature phones often used for USSD can make the platform nonintuitive to navigate for novice users and present eyesight-related accessibility problems for others. Wakalas have been shown to be effective intermediators who are already leveraged by their communities to support their use of feature phones and associated USSD applications \cite{Raghunath_2023}. The five pilot villages were randomly sampled from villages in Kagera that were not enrolled in the HH survey (to avoid confounding the collaborators' study). Between December 2022 and January 2023, authors 1, 7 and local enumerators located and enrolled two wakalas as intermediators in each village with the help of a village guide. Sampled HHs in each village were chosen through random walks that were stopped at 20 HHs. Sampled HHs were screened by asking about their comfort with USSD. Those who responded as being \textit{very comfortable} with USSD were excluded, and those remaining were included. (The protocol after sampling, including training and intermediation, is included in \ref{wakala}.) Following a one-month pilot intermediation period, authors 1, 7 and local enumerators conducted an endline phone survey with wakalas and participants, which included quantitative and qualitative feedback on the experience. 

\subsubsection{Data Analysis}
We inductively coded qualitative feedback in Atlas.ti. The codebook contained the following codes: interaction with wakalas, interaction with community members, trust, benefits of intermediation, and challenges with intermediation. We used Python to analyze quantitative responses. Further, we compared usage logs between wakala-intermediated HHs and HHs in the larger survey that were similarly (un)comfortable with USSD to identify differences in use.

\section{Results}

Below, we describe results of the HH survey (\ref{demog}), the design, development and deployment process for our apps (\ref{DDD}), the use of the dual-platform technology (\ref{use}), and the intermediation pilot (\ref{wakala}).

\subsection{Demographics and Technology Landscape}\label{demog}

We detail below the results of the HH survey, including technology use and comfort and wakala utilization in our sample.

\subsubsection{Technology Use and Comfort}
Of the 1014 HHs sampled, 7.1\% reported having a smartphone, and 92.9\% reported having a feature phone. When asked whether the HH's home was connected to the electric grid, 83.2\% said no, and 16.8\% said yes. 99.51\% of survey participants (n=1009) reported use of their phone to support their farming. The predominant use-case was to place phone calls to other farmers or extension officers to discuss planting decisions. HHs were asked if they used any USSD-based services on their phone, with M-Pesa offered as an example USSD technology. 68\% reported use of USSD, and 32\% reported non-use. Women who did not use USSD exceeded  men by 10.2 percentage points (37.7\%  vs. 27.5\%, respectively). 

\begin{figure}
    \centering
    \includegraphics[width=\linewidth]{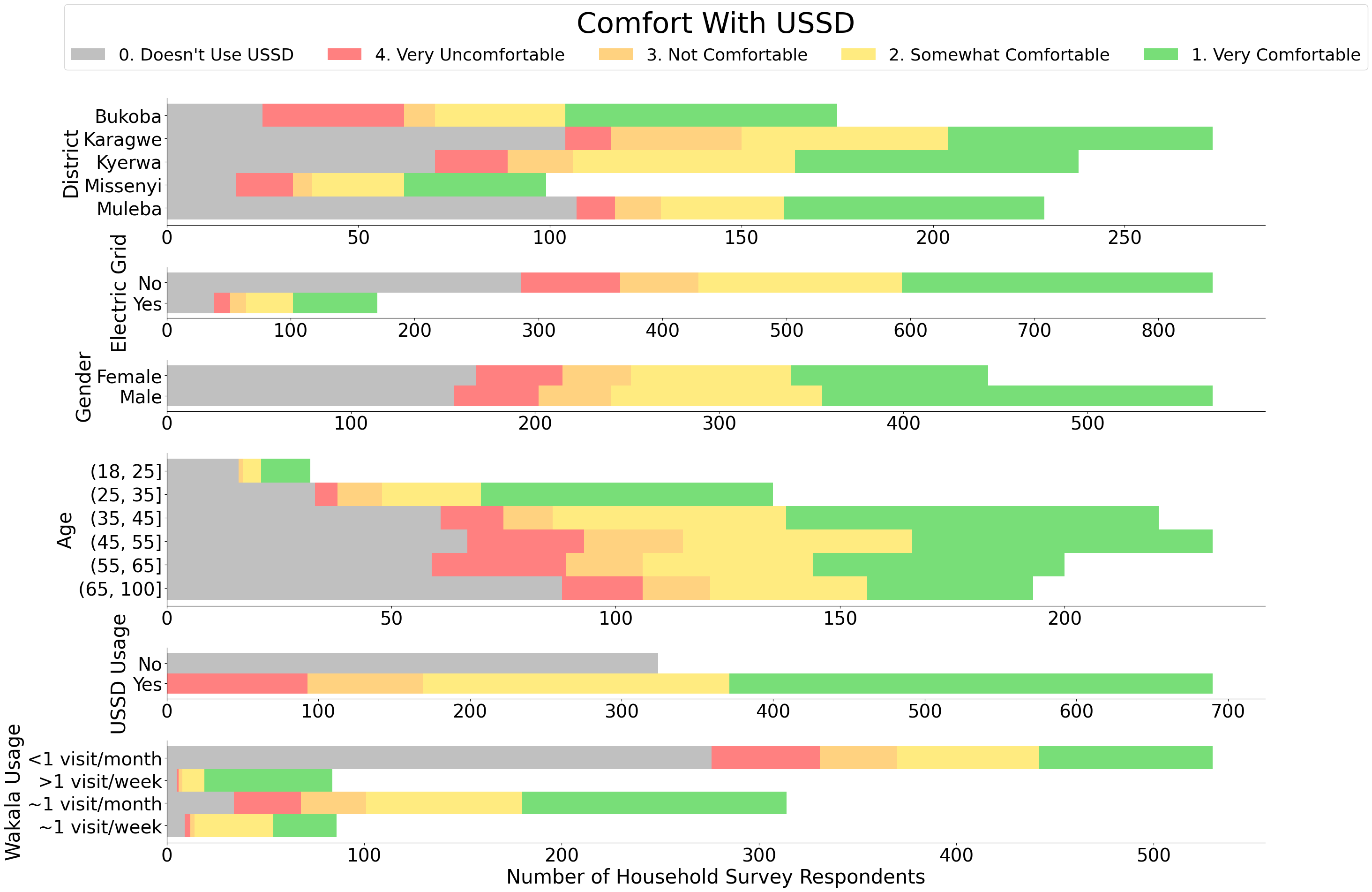}
    \caption{Demographics stratified by comfort. Women are less likely to use USSD and self select into Comfort Group A; however, men and women are equally proportioned in Comfort Group B. Older people in our sample are less likely to be users of USSD and  less comfortable on average. Over half of those in Comfort Group B visit wakalas at least once a month. District and grid connectivity are not correlated with comfort.}
    \label{fig:comfort-by-demo}
    \Description{Figure 1: "This figure includes several bar charts that break down demographic characteristics by self reported comfort with USSD. Women are less likely to use USSD and self select into Comfort Group A, however, men and women are equally proportioned in Comfort Group B. Older people in our sample are less likely to be users of USSD, and are less comfortable on average. Over half of the people in Comfort Group B visit wakalas at least once a month. District and grid connectivity are not correlated with comfort."}
\end{figure}

HHs that reported use of USSD were asked about their comfort level when using USSD-based applications, with answer options as follows: (1) \textit{Very Comfortable}, “I don’t need help," (2) \textit{Somewhat Comfortable}, “I rely on help on occasion,” (3) \textit{Somewhat Uncomfortable}, “I rely on help often,” and (4) \textit{Very Uncomfortable}, “I need help most of the time or always.” 46.2\% of HHs reported being very comfortable with USSD, whom we refer to as "Comfort Group A," and the remaining 53.8\% reported needing help on occasion (29.3\%), often (11.0\%), and always (13.5\%). We refer to these collectively as "Comfort Group B." A demographics breakdown of the HH survey participants grouped by self-perceived comfort is shown in Figure \ref{fig:comfort-by-demo}. Table \ref{tab:comfort-dist} shows the overall distribution of comfort. 

\begin{table*}[ht]
\begin{tabular}{@{}p{0.2\linewidth}p{0.3\linewidth}p{0.3\linewidth}@{}}
\toprule

\textbf{People} &\textbf{Group} & \textbf{Comfort}\\ \midrule
Non-Users: 32\% & - & -  \\
\\\midrule
Users: 68\% & - & -  \\
& A - do not need help &  Very Comfortable: 46.2\%  \\
&  B - do need help &  Somewhat Comfortable: 29.3\%  \\
& B - do need help &  Somewhat Uncomfortable: 11.0\%  \\
& B - do need help &  Very Uncomfortable: 13.5\%  \\
\bottomrule
\end{tabular}                         
\caption{\label{tab:comfort-dist} The distribution of USSD use and comfort in our sample.}
\vspace{-15pt}
\end{table*}

Self-reported comfort differed with age and gender according to an ANOVA test (P << 0.001), with younger and male gender people reporting higher comfort. However, differences in comfort were less significant for the < 40 age group, suggesting that the younger generation might be bridging the gender gap in comfort with technology (> 40 years: P = 0.020; < 40 years: P = 0.099). 

The pattern of males being more comfortable than females is statistically significant (P << 0.001); however, though men exceeded women in the Comfort Category 1 by 13.3 percentage points (37.3\% vs 24\%), women and men were proportional in categories 2-4. There are subgroups of comfort for both males and females (one group with higher comfort and one with lower). Of the collected variables, age is the most explanatory for the observed bimodality in comfort (P = 0.065). District or village were not associated with comfort (P = 0.073), indicating that perhaps there was no periurban/rural split in technology comfort.

\subsubsection{Wakala Utilization}
All HHs (including non-users of USSD) were asked how often they visited a wakala: 50.7\% said less than once a month (n=565), and the remaining 49.3\% reported at least once per month; 31.5\% said exactly once a month (n=322), 10.1\% said once each week (n=90), and 7.7\% said multiple times a week (n=86). Interestingly, HHs that report being very comfortable using USSD visit the wakala more often than those who are somewhat or very uncomfortable. This shows that comfortable USSD users have  increased engagement with services like M-Pesa, causing them to visit more frequently to deposit or withdraw cash, for example. 40\% of those self reported as needing help (somewhat comfortable, somewhat uncomfortable, or very uncomfortable) visited the wakala at least once a month. 

\textit{\textbf{Summary.} In general, we find low smartphone penetration (7\%) in this periurban and rural population. Although a large majority report using their phone for calling and texting, 30\% report non-use of USSD and 53.8\% need support using USSD. There is general heterogeneity in comfort level even though age and gender are associated with comfort. This underscores the importance of designing USSD technologies with increased usability and the challenges with selecting one subsection of the population to provide potential intermediation or support for new technologies.} 

\subsection{Design, Development and Deployment}\label{DDD}
Through iterative rounds of user testing (Section \ref{dev_and_dep}), we collected qualitative feedback and used it to inform design and implementation choices. Below, we describe our USSD and Android applications, design choices used to facilitate trust, and differences in use and preferences between the apps.

\subsubsection{USSD Application}
To access our USSD application, users dial a shortcode on their phone to initiate a session with their MNO. The MNO then requests a ``screen" from the USSD server consisting of up to 160 characters of plain text to display on the user's phone screen. While the session is active, the user can respond to each screen with text or numbers and be served a new screen in response to each input. A MySQL server is queried through the Django ORM to generate dynamic screens based on user inputs (e.g., for text searches), and these screens are then cached in memory when possible. Once the user has navigated through 6-8 screens (depending on the path they traverse), they will see a list of businesses corresponding to their search.

Through user testing, we first improved the search functionality to increase ease of navigability. When asked whether having multiple screens in a shortcode application deters their use, 46.4\% of users in our testing groups agreed. Therefore, we decided to break categories into logical subcategories where necessary, while limiting the number of sub-screens where possible. Though Weld et al.'s pilot had crops, livestock, and agricultural products \cite{weld2018ekichabi} grouped together during search, we chose to split them into individual categories to allow users to be specific about their search, eliciting the following user feedback - \textit{"Search through the menu is easy to avoid confusion on what you want, since you go direct to your need/concept using the specific choices."} Splitting into subcategories also eliminated the occurrence of "too many results" screens for many categories of searches. The final main menu screen offered 4 options: (1) \textit{filter by category}, which prompts the user to select a sector, sub-sector, geographic district, village, and subsequently sub-village, (2) \textit{filter by location}, which does the same as option 1 but starts with geographic filter options followed by sector and sub-sector filters, (3) \textit{text input search}, which lets users enter a string that corresponds to a sector, sub-sector, product, or location, bypassing the need to navigate through any USSD screens before seeing a list of relevant businesses, and  (4) a \textit{help screen}, which walks users through available search options. Filter options are suspended and jump directly to a list of matching businesses when prompted by the user or when 10 or fewer businesses match the previous filters. 

\begin{figure}
    \centering
    \includegraphics[scale=0.2]{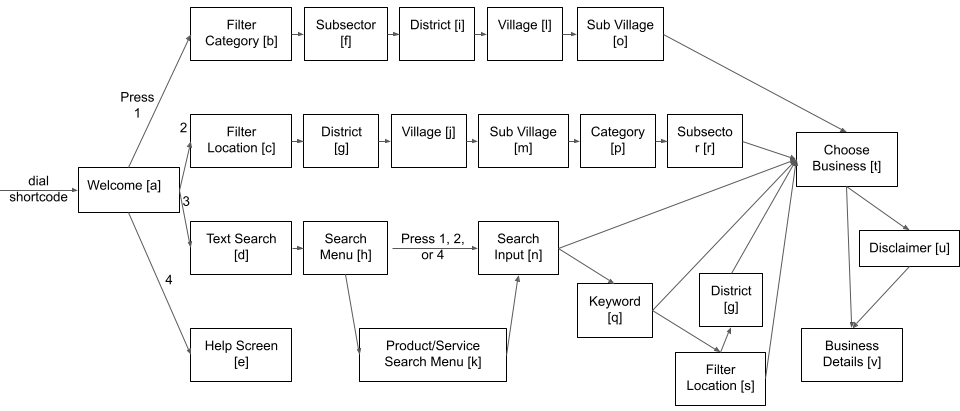}
    \caption{Users navigate through USSD screens using numerical and alphabetic inputs for text-based search.}
    \label{fig:ussd-flowchart}
    \Description{Figure 3: "This figure is a process chart of all the traversals of the USSD application's search hierarchy. In the first step, users can filter by category, location and text input, or navigate to the help screen. If filtering by category, they select subsector, district, village and subvillage. If filtering by location, they select district, village, subvillage, category, and subsector. If text search, they enter string and then can follow filter by location's steps. The help screen provides simple feedback on how to use the application. All of these traversals lead to business detail screens."}
\end{figure}

Several users expressed interest in the typing-based search to avoid narrowing their search through menus. However, spelling challenges prevented this from being as useful as anticipated - as reported during feeedback sessions: \textit{"If you mistake of some spells you may end up get wrong information, eg Korogwe vs Karagwe, Kyera vs Kyerwa."} To circumvent this problem, a "keyword" selection screen was added after text searches to let users choose from a list of close matches, reducing the occurrence of unhelpful "no matches found" screens that were prevalent in Weld et al.'s pilot \cite{weld2018ekichabi}. Figure \ref{fig:ussd-flowchart} shows the flowchart of USSD screens for the eKichabi application.

\begin{figure}
    \centering
    \begin{subfigure}[b]{0.2\textwidth}
        \centering
        \includegraphics[width=\linewidth]{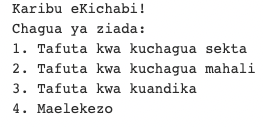}
        \caption{Welcome Screen}
    \end{subfigure}
    \begin{subfigure}[b]{0.2\textwidth}
        \centering
        \includegraphics[width=\linewidth]{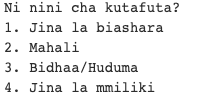}
        \caption{Text Search Menu}
    \end{subfigure}
    \begin{subfigure}[b]{0.2\textwidth}
        \centering
        \includegraphics[width=\linewidth]{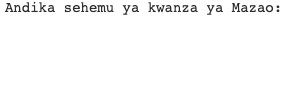}
        \caption{Search Input}
    \end{subfigure}
    \begin{subfigure}[b]{0.2\textwidth}
        \centering
        \includegraphics[width=\linewidth]{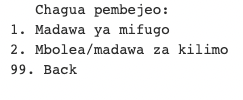}
        \caption{Keyword Select}
    \end{subfigure}
    \begin{subfigure}[b]{0.2\textwidth}
        \centering
        \includegraphics[width=\linewidth]{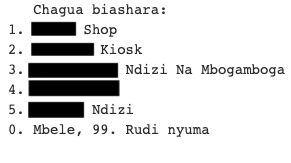}
        \caption{Business Select}
    \end{subfigure}
    \begin{subfigure}[b]{0.2\textwidth}
        \centering
        \includegraphics[width=\linewidth]{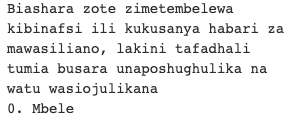}
        \caption{Disclaimer}
    \end{subfigure}
    \begin{subfigure}[b]{0.2\textwidth}
        \centering
        \includegraphics[width=\linewidth]{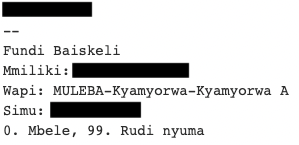}
        \caption{Business Detail Screen}
    \end{subfigure}
    \caption{A few screens from the USSD application demonstrating what a first time text search might look like; phone numbers and names have been obscured to preserve privacy. A: The first screen displays the 4 menu options -- category search, location search, text search, and help. B: After selecting the 3rd option, the user has 4 text search options -- by business name, location, products/services, or owner's name. C: The user is prompted to spell out part of the thing they are looking for. D: If there are many matches, or no matches for the provided spelling, the user is prompted to select the keyword that most closely matches what they were looking for. E: After selecting, the user will either be prompted to further filter by location, or in this case since there were only 5 results, be prompted to select businesses to look at directly. F: The first time a business is selected, the disclaimer will be displayed. G: Finally a detailed overview of the Business name, economic sector, owner name, location, and a contact phone number is displayed.}
    \label{fig:ussdscreens}
    \Description{Figure 2. The USSD screens are shown. They are each 4-8 line screens, displaying plain text menus. The text is in Swahili. There are "99. Back" options once the users have gotten far enough into the search hierarchy.}
\end{figure}

User observations underscored that fast application performance was necessary to avoid MNO-implemented timeouts that reset users to the first screen - as one user said: \textit{"Some search took so long to display, I had to start from the beginning"}. This presented a steep technical challenge since our database grew from 500 to around 10,000 businesses. To address this, we performed database optimizations, including: implementing a Redis cache for session information, changing the session storage format from pickles to compact strings, adding lazy loading for search results to reduce computational overhead, caching pre-computed screens, adding proper database indexing for fuzzy search, and using MySQL instead of SQLite to handle concurrent connections. Collectively, these significantly boosted performance, with request times dropping from 5+ minutes of server computation when using Weld et al.'s implementation \cite{weld2018ekichabi} to 100-800 ms using our new system. This signified an average performance improvement across a single session of more than 7000x as indicated by benchmark and stress tests.

\subsubsection{Android Smartphone Application}
\begin{figure}
    \centering
    \begin{subfigure}[b]{0.2\textwidth}
        \centering
        \includegraphics[width=\linewidth]{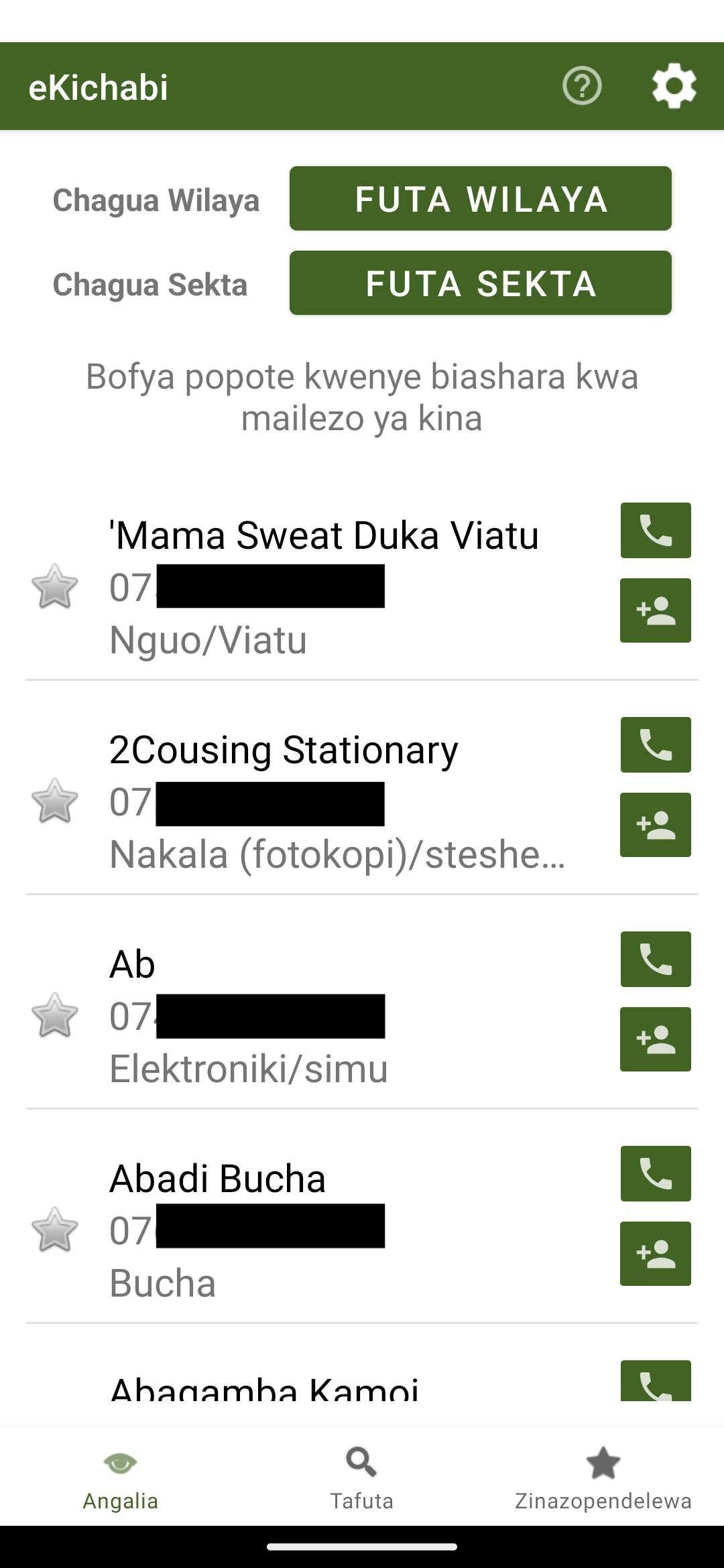}
        \caption{Initial Screen}
        \label{fig:scroll}
    \end{subfigure}
    \begin{subfigure}[b]{0.2\textwidth}
        \centering
        \includegraphics[width=\linewidth]{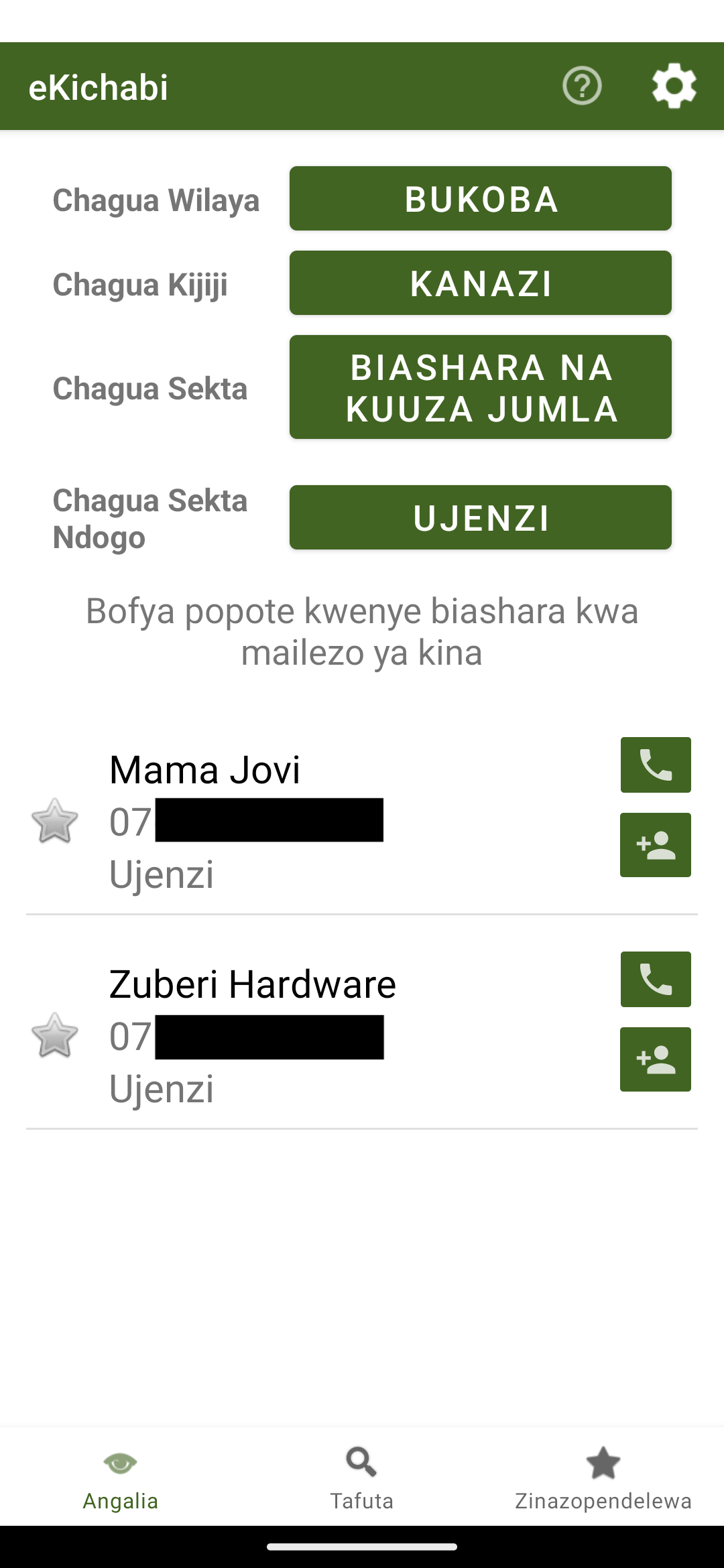}
        \caption{Filtering by Menu}
        \label{fig:filter}
    \end{subfigure}
    \begin{subfigure}[b]{0.2\textwidth}
        \centering
        \includegraphics[width=\linewidth]{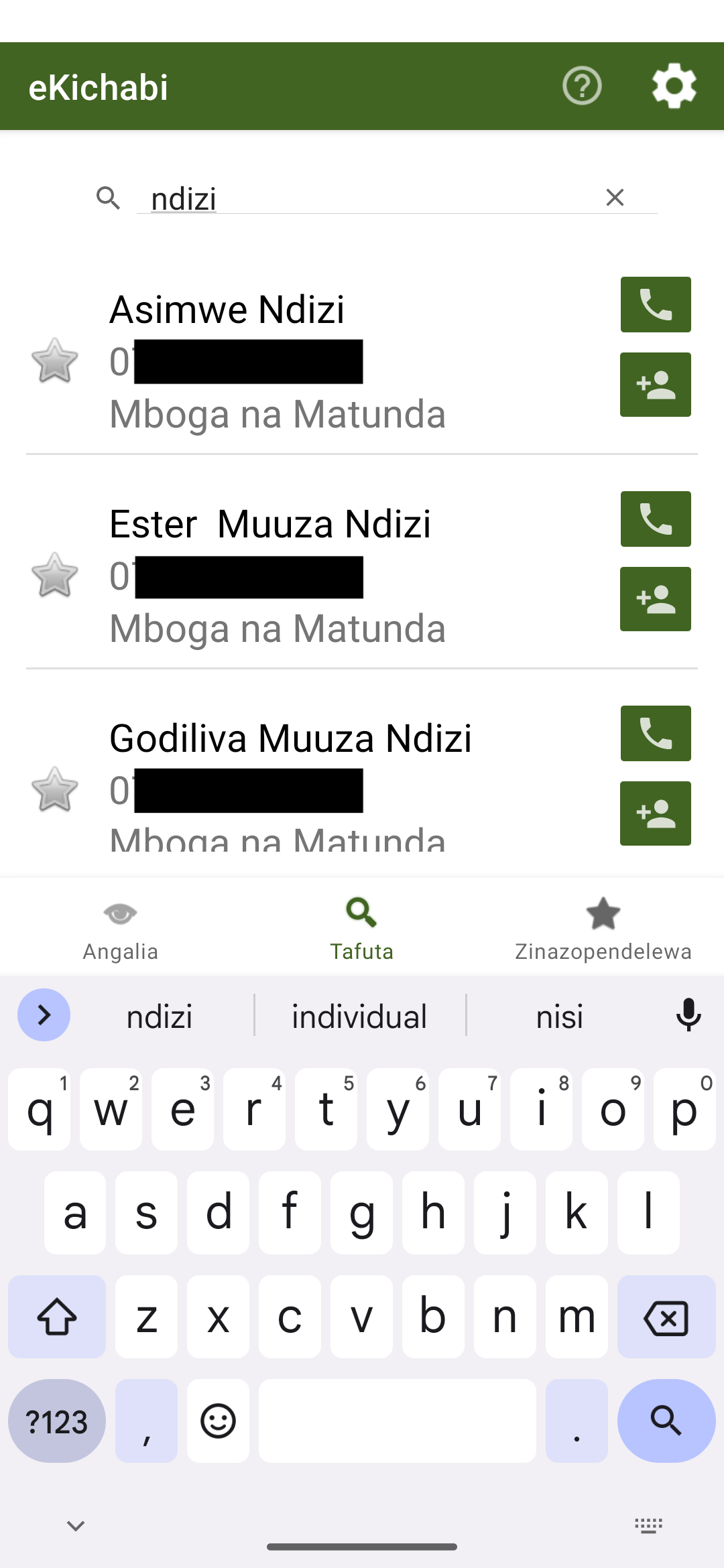}
        \caption{Filtering by Text Search}
        \label{fig:search}
    \end{subfigure}
    \begin{subfigure}[b]{0.2\textwidth}
        \centering
        \includegraphics[width=\linewidth]{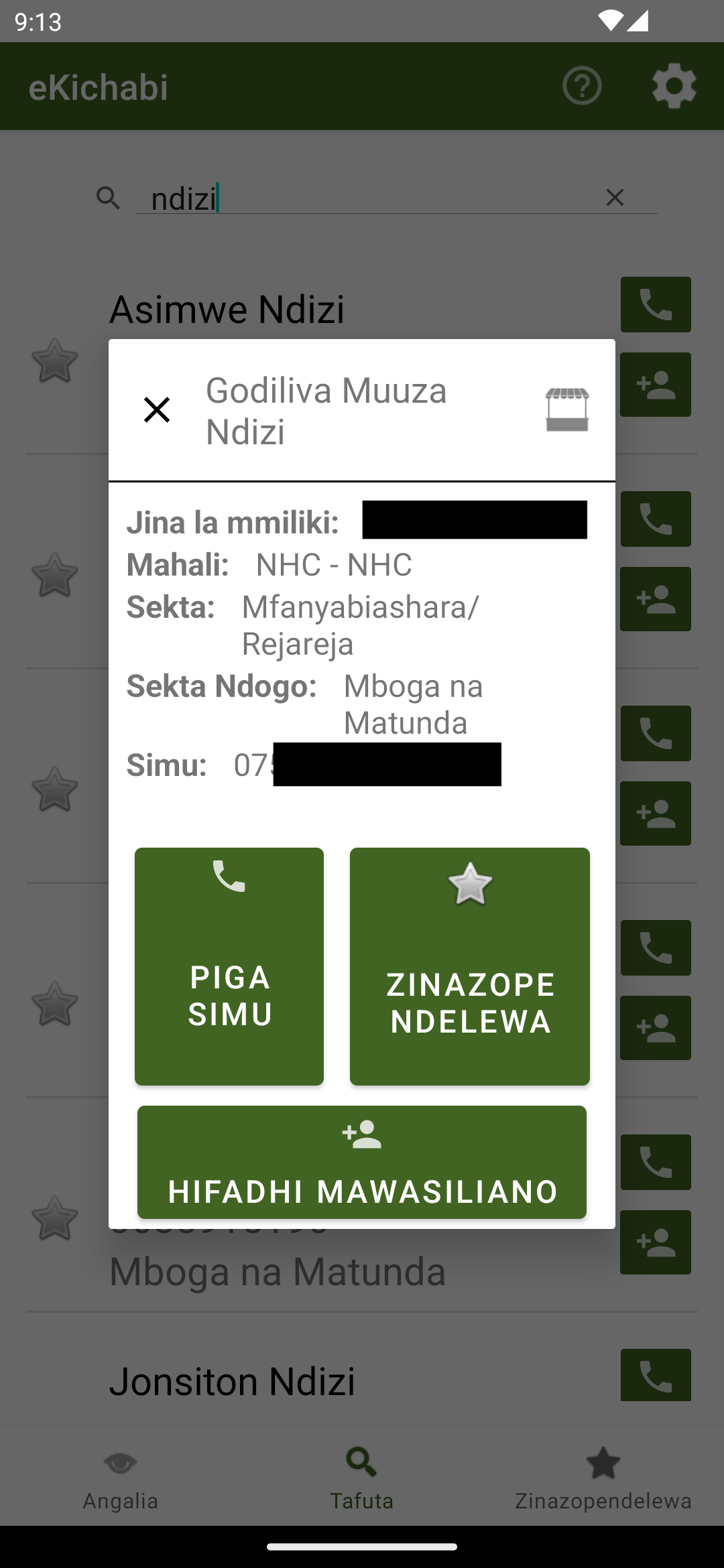}
        \caption{Business Detail Screen}
        \label{fig:business_detail}
    \end{subfigure}
    \caption{A few screens from the Android application; phone numbers and names have been obscured to preserve privacy. A: The user can scroll through the businesses, favoriting, adding to contacts, calling and viewing more details, as desired. B: The user can filter the businesses by district, village, sector and sub-sector to make results more relevant. C: The user can find businesses via text search: a scrollable list of businesses dynamically appears as the search is being narrowed. D: The user can click on a business name to open a screen showing business name, owner name, phone number, location, sector, and sub-sector.}
    \label{fig:android}
    \Description{Figure 4: "This figure contains 4 screenshots of the Android app with phone numbers and names have been obscured to preserve privacy. Screen A: The user can scroll through the list of businesses upon opening the app, favoriting, adding to contacts, calling and viewing for more details, as desired. Screen B: The user can filter the businesses by district, village, sector and sub-sector to make results more relevant. They do so by selecting the menu buttons at the top of the screen and choosing an item from the dropdown menus thereafter. Screen C: The user can find businesses via text search through a bar at the top of the screen that accepts typed input. a scrollable list of businesses dynamically appears as the search is being narrowed. Screen D: The user can click on a business name to open a pop up screen showing business name, owner name, phone number, location, sector, and sub-sector. This screen has buttons that allow users to call the business, favorite the business, or add the business to contacts.}
\end{figure}

After users download the Android app, which we developed as a GUI-based directory, they are shown a short tutorial followed by a list of all businesses in the directory, through which they can freely scroll (Figure \ref{fig:scroll}). The app provides filtering options based on geographical locations at district, village, and sub-village levels, as well as sectors and sub-sectors (Figure \ref{fig:filter}). Text search dynamically displays relevant results after the user inputs their query (Figure \ref{fig:search}). Additionally, the GUI features convenient buttons for actions, e.g., accessing the list of favoriting businesses, calling businesses directly, and adding businesses to the user's contacts (Figure \ref{fig:business_detail}). We made many improvements over the course of Android user testing. These include optimizing images and assets for faster loading, adjusting colors to work well in bright light, creating a more robust help screen, adding animated GIFs to explain features, and general bug-fixing. The Android app relied on the same server and database as USSD, centralizing the data and making it easier to keep it current as well as reducing costs associated with scaling by removing the need to maintain two separate server instances. 

The main challenge encountered in Android app development was making it work without an internet connection. We chose this as a design principle because most users reported having an internet connection in the village center but most often not in their homes. We considered several solutions, one of which had the Android app use a USSD connection when no internet was available and augmented data appearance through the GUI, but this was deemed too unreliable for deployment due to (1) limitations in different Android operating systems' support of USSD connections, and (2) the USSD gateway's inability to accept initial payloads upon starting USSD connections. The final design kept a local copy of the phone directory (1.8 Mb). Any subsequent directory updates are efficiently performed when the device has an internet connection: an endpoint was created to check if the directory is up-to-date so it need not update its local copy unless necessary. Internet connection is required only for the initial app download and user authorization. 

Our final app was 6.73 Mb of data, which is 14 MB smaller than the median for this type of application on the Play Store, as displayed on our developer console. We stored usage logs in a custom binary format for compactness because offline storage space and online data transfer were limited resources. Logs returned to our server are limited to 100 KB local storage before uploading, which is sufficient for 6250-20000 actions each given our custom formats. In 2023, 6.73Mb of data cost \~67 TSh or \$0.02 USD.

Both the USSD and Android apps use \textit{whitelisting} (i.e., allowing only the phone numbers in our study to access the intervention) to enforce specific treatment groups and villages as needed by the collaborators. The whitelisting mechanism is based on a common server endpoint that standardizes the phone number format to account for irregularities in phone number data and then compares against a hashset of pre-approved, formatted phone numbers. One challenge in authenticating users for the Android whitelist was the need to collect phone numbers from users to grant them access to the service. This arose from the technical limitations of the previous versions of the Android SDK and older hardware, which were common in this context; for example, dual SIM implementations, older SIM security systems, and lack of support on old Android operating systems created challenges when accessing phone numbers programmatically. To maintain compatibility with the hardware commonly found in developing regions, we decided to treat phone numbers as one-time passwords rather than as 100\% reliable identity providers through hardware. Although other solutions, such as sending one-time access links, were considered, they were eventually discarded since they did not offer significant security benefits and would have posed a barrier to use for less tech-literate demographics. We settled for allowing the user to input their own phone numbers, trusting that they would not share them with others who were not whitelisted. Analysis of the Android logs showed no anomalies, suggesting that login with one's phone number was not an authentication problem.

\subsubsection{Trust}
Trust was a significant factor in acceptance and use of the app and arose repeatedly in user testing. As researchers have found, text message or USSD-based scams are becoming commonplace in this area \cite{Raghunath_2023}. Several users questioned why they should trust the phone numbers in the directory, what recourse they had if they were conned by a contact they made through the directory, whether the app would validate the businesses, and so on. One user asked: \textit{"What if I call a transporter through the directory and he comes and kills me?"} We made three changes to the app training protocol and the technology to inspire trusting interactions, as follows. 

\begin{enumerate}
    \item \textbf{Users were skeptical about our incentive when creating and disseminating the directory. In addition, they wondered how we selected businesses to include in the directory.} In our training materials, we explained our research study and goals and mentioned the prospect of handing over a similar project to the government of Tanzania. We also made clear that the information in the directory was assembled by our research team through a survey that we conducted when we worked with the support of local village extension officers (VEOs) to collect information. Every business listed in the directory was visited by one of our team members to gather information in person. 
    
    \item \textbf{Users were accustomed to building trust by meeting face-to-face. The new potential for interactions inspired questions as to whether we guaranteed transactions as part of our role in maintaining the service.} We modified the training to encourage continuation of face-to-face meetings even when people make initial contact through eKichabi. We added a disclaimer the first time either app was used, alerting users that they should take care when contacting individuals and that we did not guarantee transactions: \textit{All businesses have been visited in person to collect contact information, but please use discretion when transacting with unknown individuals. It is possible that some businesses will close this year or change their business practices. But we hope and expect that the directory will remain very relevant for years to come}. User feedback gave us other ways to improve this mechanism in later iterations of this study. For example, one user said \textit{"You can also think of adding village officials to each contact listed on the last screen in any case one wants to contact the official person for verification of the business since those officials knows well people around their area of administration."}

    \item \textbf{Users would trust the directory if national tax IDs were collected from businesses and added to the directory} Until user testing, we listed only business name, contact information, and sector in the directory. Understandably, not all businesses are registered in Tanzania, and national IDs are too sensitive to include in the survey or directory. Once we understood that Tanzanians do not typically identify businesses by their names but rather consider the business owner as representative of the business, we added owner names to the business information we provided in the directory. We created an associated search pathway to allow searches through owners. Although this modification added technical complexity, user engagement grew during testing sessions when users saw their friend's or family's business in the directory, and the sense of legitimacy granted by these interactions was extended by users to other businesses listed in the directory.
   
\end{enumerate}

\subsubsection{Prominent Differences and User Preferences}

Though we designed both USSD and Android apps to present the same service across different platforms in similar ways (filtering, text search, etc.), there were unique and inherent advantages of each, as listed in \ref{tab:diffs_in_apps}. 

\begin{table*}[ht]
\begin{tabular}{@{}p{0.2\linewidth}p{0.4\linewidth}p{0.4\linewidth}@{}}
\toprule
\label{table:mytable}

\textbf{Theme} & \textbf{USSD App} & \textbf{Android App}\\ \midrule
Navigation & Users navigate through USSD screens with numerical input corresponding to selections. To go to the previous screen, they enter 99; to go to the next screen, they enter 0. & The Android app leverages the smartphone touchscreen to move back. `Next' is not a necessary function since all relevant businesses can be scrolled through at once. \\\hline

Visibility of Businesses & Users must navigate a minimum of 6 screens before businesses become visible and open business details screens individually to see more than the business name. & The Android app displays the list of unsorted businesses as soon as the user opens the app (see \ref{fig:scroll}), with business name, owner phone number, and economic sector/product type available without clicking on the business details screen.\\\hline

Search & Screen size and character limits limits put a ceiling on the number of business names displayable on a single screen, i.e., typically 5 to 6 businesses. This means there can be tens of pages of responses for some queries that the user must traverse sequentially. & Android search is dynamic, and the list of businesses updates as the search is being narrowed. As many businesses as are applicable appear during the search.\\\hline

Additional functions & Users must physically write down contact numbers from the final business screen since feature phones do not allow text copying and pasting. Use of the help screen requires returning to the first page of the app.& Android users can favorite businesses (which then appear on a separate tab), save phone numbers to contacts, and call businesses through the app. The help button can be toggled from any screen. \\\hline
Costs & No costs are incurred when users dial the USSD app. & Android users exhibit reluctance to download the app, citing data costs. \\\hline

Authentication & The USSD app automatically authenticates users with their phone number since it has access to that information from MNOs. & Operating system version challenges on popular smartphones cause users to self authenticate as opposed to using a hardware-based guarantee from sims.\\\hline

User Experience & USSD has strict character limits of 160, and on feature phones, it is often necessary to scroll down to display a single screen. This makes USSD more laborious to navigate. Further, feature phones have no way to increase screen brightness or increase text size. & The Android app displays more data while remaining readable and having a more intuitive user interface. Further, smartphones can increase screen brightness and text size to suit users and the environment. \\\hline

Potential Interruptions & If the telecommunications network is going in/out, the USSD connection fails, and the user must resume from the beginning when the connection returns.& Once the app is downloaded, there is no connection necessary unless the directory changes and new businesses must be downloaded (even this can be done whenever WIFI becomes available, and the app remains  usable in the interim). \\\hline
Trust & Trust building comes in the form of a disclaimer and the addition of business owner names to the directory. There is no easy way to corroborate information validity given the number of steps needed to find a business.& Trust is perceived to be greater: although a similar disclaimer is presented, owner names are visible with the list of businesses when the app is opened, and familiar businesses can be located with ease.\\\hline
\end{tabular}                         
\caption{\label{tab:diffs_in_apps} Differences between the USSD and Android apps.}
\vspace{-15pt}
\end{table*}

\paragraph{Preferences}
During all user testing sessions, our USSD and Android apps were tested by smartphone owners (SPOs), while USSD alone was tested by feature phone owners (FPOs). When we used smartphones with beta versions of the Android app already downloaded to SPOs for trial purposes, the Android app fared much better than USSD across groups (76.9\% of SPOs in these groups preferred Android over the USSD app). Preferences for the smartphone app were magnified when there were network connectivity issues impacting USSD since the app works completely offline. Confusion with USSD concepts, such as selecting "next" and "back" numbers to navigate between screens, further enhanced the preference for the more intuitive smartphone app navigation. 

However, when we asked SPO groups to download the app onto their own phones, we witnessed significant reluctance, which limited their ability to experience the app's additional features. SPOs explained that data on Androids was harder to gain access to and that they were accustomed to using USSD, which they understood to be free. Therefore, \textit{there was an overall preference for USSD by both smartphone and feature phone users.} Interestingly, both FPOs and SPOs preferred using the USSD app on a smartphone compared to a feature phone, citing increased usability. Their main rationale included complaints about eyesight and text readability on feature phones. The USSD screens of \~160 characters took multiple pages (and therefore required scrolling) on the small displays of feature phones but appeared on one page on smartphones, with displays that could be pinched to increase font size. Therefore, the affordances of the smartphone impacted the USSD application positively.

From the perspective of trust and usefulness, more SPOs reported that the Android app appeared trustworthy and useful. This might be due to faith in the new smartphone technology they have already purchased or to the immediate visibility of businesses and owner names without much upfront navigation, as is required by the USSD app. In final rounds of user testing, participants offered the following benefits of our app used on their platform of choice:

\begin{enumerate}
    \item It makes it easier to sell and buy when you know about the area businesses. No middle men are needed; you can directly contact the supplier/buyer and so increase profit when selling and reduce cost when buying.
    \item You can contact the business to gain an assurance about goods and service availability before making an order or visiting a buyer/supplier.
    \item For smaller business and service providers, especially in the local area, it is an easy way to access larger business buyers/suppliers that are always not found in the local area.
    \item It can be used for service provider job recruitment or to display sales information  because qualifications or notes can be added to the business detail screen.
\end{enumerate}

\subsection{Usage}\label{use}
Below, we report on use of the dual-platform intervention.

\subsubsection{USSD Use and Comfort}
The USSD application received 122,000 actions over 2939 sessions, with 266 unique HH users of the 1014 whitelisted households. Usage statistics---such as session duration, number of sessions, retention, use of back buttons, and text search---were highly correlated with number of businesses visited (p << 0.001). The final business screens, which offer the payload of information that the service aims to provide, can thus be seen as tied to the value provided by the service to the user; 782 business screen were visited over the uptime. 

The overall mean number of businesses visited by males was 1.046 (s.d. = 6.128), which breaks into two distinct groups, as described below. The first had a mean of 0.27 business visited (s.d. = 0.98), and the second had a mean of 25.17 businesses (s.d. = 24.60). The female group fell between these, with a mean of 0.38 business visited (s.d. = 1.68). ANOVA analysis shows lack of significance in number of businesses visited between genders (P = 0.086); thus, both men and women navigated searches and achieved a similar amount of value from the application. 

The only significant predictor of number of businesses visited is user age (P = 0.006). Males were generally younger in our sample, with a mode age group of 35-45 years compared to females, who were between 65-100. Interestingly, the bi-modality within the male group is explained by age since the group with more business visits had a lower average age of 39 years compared to that with fewer visits (average age 49). The pattern that younger people visited more businesses held for both genders (All: P = 0.006; Male: P = 0.014; Female: P = 0.026). 

The number of businesses visited did not differ across districts (P = 0.937), indicating that there was no peri-urban/rural split in relation to use. However, the number differed significantly across whitelist dates (P = 0.012), indicating that seasonal farming practices have a large impact on use of agricultural information dissemination systems.

Notably, the number of businesses visited did not differ across comfort levels (P = 0.556). However, Comfort Groups A and B show interesting usage differences (see summary in Table \ref{tab:ussd_use_comfort}). For example, all Android application users came from Comfort Group A. In addition, Comfort Group A had a larger number of sessions, made greater use of the back code (0), ended up on more final business pages, and had longer session durations through ANOVA analysis, with a significance level of 1\% .

\subsubsection{Android Application}
Android use peaked at 42 active devices. Of the 42 downloads, 40 unique users accessed the application. Android logs tracked 7 different action types: favoriting/unfavoriting businesses, calling a business from within the application, adding a phone number as a contact, opening the business detail screen of a business, searching with filters, and searching with text. Favoriting was used by 12.5\% of Android users, with 40 favorite/unfavorite actions. There were also 51 call actions from within the app. Similarly, 15 businesses were added to phone contacts. When stratified, the user population was not sufficient to observe significant differences in use across comfort or demographics.

\begin{table*}[ht]
\begin{tabular}{@{}P{0.33\linewidth}P{0.33\linewidth}P{0.33\linewidth}@{}}
\toprule
\textbf{Contributing Percentage} & \textbf{from Comfort Group A} & \textbf{from Comfort Group B} \\ \midrule
eKichabi USSD users & 21\%  & 12\% \\
eKichabi Android users & 0.3\%  & 0.0\% \\
Total Sessions & 488 & 205 \\
Average sessions per user & 7 & 5 \\\hline
\end{tabular}                         
\caption{\label{tab:ussd_use_comfort} Differences in use by comfort level.}
\vspace{-15pt}
\end{table*}

\subsubsection{Comparison of USSD and Android App Use}\label{comparison_use}
Android app use was more complex and fostered higher user retention than the USSD app. There were 22 actions per user on Android (874 total) compared to 42 USSD inputs per USSD user who used the app at least once. Though this appears to indicate greater engagement on USSD, each USSD input confers less "valuable" information, with 6-8 actions needed before viewing the business screen; the Android homepage presents all businesses for the user to scroll through before any actions are taken (see \ref{tab:diffs_in_apps}). In addition, there was an average of 5 views of business detail screen actions per user (193 total) on Android compared to an average of 3 open business screen actions for USSD (1785 total). This suggests that Android users were keener to know more about the businesses they browsed through or had easier access to businesses, encouraging increased exploration. 

Another significant difference between USSD and Android was in filter actions: Android had 462 browse actions (about 12 per user) compared to USSD's 0.2 browse by category or location actions (126 in total). Thus, filtering through businesses was substantially easier and more frequently done on Android. Likewise, Android had 85 text search actions (about 2 per user) compared to USSD, which had less than 0.5 text search actions per user (288 in total). Further, Android users returned to the app on an average of 2.65 different dates compared to each USSD users' 1.72 different dates, suggesting that the Android app may have been more helpful to its users. 

\subsection{Wakala Intermediation Pilot Study}\label{wakala}
Our small scale pilot study aimed to assess the utility of intermediation in promoting USSD app acceptance and use in users from Comfort Group B. We enrolled 10 wakalas and 31 wakala pilot HHs (pHHs) across the five villages in the study. A village guide sampled wakalas and enquired about availability and interest in helping with intermediation before Authors 1, 7 and enumerators arrived at the villages to conduct training. Table \ref{tab:wakala_dist} shows the distribution of study participants. 

\begin{table*}[ht]
\begin{tabular}{@{}P{0.33\linewidth}P{0.33\linewidth}P{0.33\linewidth}@{}}
\toprule
\textbf{District - Village} & \textbf{Wakalas Enrolled} & \textbf{HHs Enrolled} \\ \midrule
Bukoba Rural - Kanazi &  2 & 8 \\
Bukoba Rural - Irogelo & 2  & 6 \\
Bukoba Rural - Izigo & 2  & 8 \\
Muleba - Rwagati & 2 & 3 \\
Missenyi - Katendaguro & 2 & 6 \\
\hline
\end{tabular}                         
\caption{\label{tab:wakala_dist} Distribution of wakalas and HHs in the pilot study.}
\vspace{-15pt}
\end{table*}

The protocol for the pilot began with the enumerator individually training each pHH and wakala on the USSD app. The same protocol was used as in the larger deployment described above. In addition, the pHHs were told that a wakala would contact them to help them use the USSD app if they were interested. The wakalas were given phone numbers of the pHHs under their purview. They were asked to text the pHHs once to offer help and send a reminder text sometime over the subsequent 3 weeks. Wakalas were asked to keep count of people who asked for help, and what kind of help. They were compensated for an estimated 2 hours of work per week across 4 weeks at their average pay, amounting to 10,000 Tsh (\~\$4 USD)/wakala for the pilot.

\subsubsection{Phone-based Endline Results}
Wakala adherence to offering their intermediation to pHHs by sending the first text was high at 90\%, but this number dropped to 50\% for sending the reminder text. Despite lower adherence to the second text, 27\% of the pHHs met wakalas in person for assistance, 18\% spoke to wakalas via phone, and 9.1\% used text messages. Of those pHHs that did not reach out to wakalas for help, some said they could use the app on their own and some tried by themselves so that they could learn without others’ help. Those that reported needing help were not from one demographic, i.e., both young/old women and men accepted wakalas' help in using the app.

Seventy percent of the walakas were asked to clarify how to use the app or for more information regarding its trustworthiness. For example, wakalas were asked how to: navigate through different categories of service; login to the system; get the USSD code when users did not remember; scroll when finding names since some HH members could not see well, and more. Our qualitative results indicate that fostering trust met with uneven success: some pHHs extended their trust in the wakalas to trust in our tool, i.e., wakalas lent legitimacy to the ICT. For example, wakalas reported that pHHs asked them about the origins of the intervention, how to contact the app owner for more information, etc. One wakala reported that two people not enrolled in the pilot came and asked for information on how to access the app and how it could help their farming. However, one wakala reported facing some resistance, saying \textit{"It is hard to make people understand the program because people think that we are thieves."}  Despite some mistrust, several users shared that they perceived the app to be useful, in particular to the business community as well as those who need services from the community. There were some challenges that the wakalas could not help users overcome, for example, one user mentioned \textit{"We need more stable network connection in our community to use the service."}

\subsubsection{Quantitative Exploration}
This pilot explored intermediation of our app through wakalas on a small scale. Therefore, it was not intended to measure statistical significance. Below, we present findings for the pHHs and Comfort Group B households (bHHs) as context to our qualitative findings, supporting further study into the impacts of intermediation on technology use in low-comfort populations.

By many metrics, pHHs used the tool more than the bHHs, and use by pHHs was more complex. For example, the proportion of unique users significantly differed in each group, with 44 unique users (12\%) from the 371 bHHs and 10 unique users (33\%) from the 31 pHHs. On average, pHHs had significantly higher engagement per user: 371 bHHs engaged in 205 sessions, and 31 pHHs had 38 sessions. pHHs also had more sessions per whitelisted number, indicating a higher retention rate than bHHs. pHHs had significantly longer session durations (175 seconds on average, compared to 30 seconds on average in bHHs), with more sessions using the back button option. Most importantly, pHHs visited significantly more businesses, i.e., had more successful traversal through the search tree (while bHHs visited 2+ business screens in less than 20\% of sessions, this was 60\% in pHHs). Both groups had a comparable number of text and category-based searches. Use of the exit and home functionality and advanced search behavior (like additional filter by location) was low \& insignificant in both groups. 

\section{Discussion}


Our paper enriches the current body of literature by advancing the understanding of digital information dissemination at scale in SSA. Here, we contextualize key findings from our work and provide design recommendations for the HCI community.

\subsection{The Future of Agricultural Information Services on Dual-Platform Technologies}

As agricultural challenges burgeon, ICTs services will need to encompass more information in a user-friendly manner (i.e., increase their application scale) as well as cater to a progressively more diverse array of users (i.e., increase their deployment scale). Our work shows that there is significant scope to create dual-platform technologies to disseminate agricultural information in SSA. Our HH survey results indicate limited smartphone penetration in the periurban and rural areas in Tanzania (\textasciitilde 7\%). The benefits of additional smartphone-related affordances seen in the HCI literature \cite{harrison2013usability, punchoojit2017usability} were reflected in our own user tests and development process (Table \ref{tab:diffs_in_apps}) and in overall usage logs (Section \ref{comparison_use}). \textasciitilde 76\% of SPOs preferred the Android app during testing and accessed more business information with less effort. Eventually, this enabled more complex use and greater retention. 

When creating the smartphone component of any dual-platform technology in SSA, designers must overcome several challenges. Although barriers to accessing smartphones are slowly being circumvented, it is important to consider the prevalent limitations of cheaper, popular smartphones, such as counterfeit phones, old Android versions, and the inability to download new apps from the Play Store, among others \cite{dodson2013minding, wyche2015real, boateng2021suacode}. If app developers and researchers are not designing around these factors, nonuse will be the default among smartphone users. Designers must also create dissemination mechanisms and incentives for installing Android apps to offset data bundle expenses, especially for tentative SPOs that stand to benefit greatly from the eventual information gain. Although we spent significant time making our Android app compact, SPOs remained reluctant to download it from the Play Store. 50.0\% of SPOs opted to use the USSD app exclusively or in tandem with the Android app, making the case for continuing to support feature phone apps. \textit{This result shows that even new SPOs who consume information through a smartphone do not necessarily change the means by which they do so.} Work on popularizing smartphone apps for financial services similarly reveals that the appetite for Android apps is lower than for USSD apps among SPOs \cite{bryan2021back, Saini_2017}. Collectively, these points raise the question of when, if at all, smartphone penetration, capabilities, related infrastructure, and user preferences will grow to sufficient mass for marginalized populations to benefit from the technology transition. It is clear that we are not at this point in Kagera. 

While USSD is a viable and widely used technology in northwest Tanzania, 30\% of the HHs we surveyed self-identify as non-users of USSD, and 53\% of users need help with USSD technologies at least on occasion. Our findings linking age and gender to USSD use are relevant to designers as a majority of farmers in Tanzania are women, who we find are less likely to use USSD. Although the literature recommends USSD for this region, scaling a USSD app to house a larger database creates engineering and design challenges that need to be addressed \cite{wyche2012dead, silver2018internet, weld2018ekichabi}. One challenge specific to information dissemination is designing short, yet informative screens for filtering through a growing database. Weld et al. \cite{weld2018ekichabi} implemented a “too many results” screen if a query resulted in more than 100 businesses, asking users to narrow their searches by answering questions on additional screens. \textit{However, if a large number of screens are an inhibitor of use (affirmed by users in our testing groups) and there are characters limits per screen (and therefore limits on businesses visible at a time), what is the marginal utility of scaling databases underlying a USSD application?} Analysis of our usage logs shows that getting to the business screens is challenging for users: only 782 business screens were visited by the 266 users that tried the service. Although access to more businesses in harder to reach areas has shown to be productive for farmers, it is hard to justify listing and maintaining an even bigger set of businesses if so few are feasibly arrived at using the USSD app. This makes us question how useful USSD is for information dissemination \textit{at scale}. 

\subsection{Intermediation as a Means to Confer Trust and Bridge the Affordance Gap}

Questions of trust (e.g., how to trust, whom to trust) were commonly raised by users, especially FPOs. Defensive attitudes about new services are common in this rural agriculture focused population and have been described previously. There are valid reasons to be cautious, e.g., several cybercrime studies mention the rising phone-based scams in this area \cite{baur2019cyber, longe2009criminal}. This problem was also raised by work in Tanzania with wakalas, where community members mention identity theft, among other challenges, when using new USSD codes \cite{Raghunath_2023}. Fostering trust through design has inherent limits that technologists must contend with. For example, we added owner names to our directory to improve user trust; however, people reluctant to try the USSD app would not learn about such trust-conferring information unless they navigated through the search tree to a business page. Further, names instilled trust best when USSD users landed on business owners that they already knew in person. In contrast, several SPOs were immediately trustful of the Android app, and we posit that this may be due to fact that users could readily search for and find known businesses and therefore easily assess the validity of the data provided. Limitations with instilling trust in USSD applications are amplified by context: official business names are largely non-existent in rural Tanzania, and enrollment of businesses to tax authorities is inconsistent. Therefore, there is no way to verify that businesses are legitimate other than to visit them in person. This puts the onus on designers to create new methods to build trust in feature phone applications for users who may be growing more skeptical of technology over time.

One way to circumvent trust and usability challenges is to create intermediation structures in the hope that users will extend trust to new tools through individuals who are locally situated and already trusted. Our pilot, similar to the integration of pharmacy workers in Bangladesh to intermediate for patients \cite{rahman2021dakter}, showed that building intermediation for dual-platform technologies has strong potential to bridge the gap caused by the lack of affordances in feature phones. We were successful in our choice of intermediary since wakalas learned the application quickly, were open to the opportunity, had a broad technical competence that allowed them to search quickly on behalf of other users, and were frequently visited by the community. \textit{Given feedback from wakalas that HHs not involved in the pilot asked them for information and how to use the service, wakalas have untapped potential for publicizing technologies that may otherwise have low uptake.} We were unable to examine their marketing support due to the constraints of the broader study we worked within. 

One prominent consideration for designers creating structured intermediation is how to select users to enroll. Our pilot showed that wakala interest and motivation was not uniform, reflecting the importance of incentivizing intermediaries \cite{moitra2016design, nicholson2019exploring} and finding a targeted user population who would be interested and benefit most from support to avoid overburdening wakalas. The literature is sparse on methods to carefully select users to support, so we considered several screening questions or demographic factors before settling on self-reported comfort. For example, (frequent or infrequent) wakala visitation could not be directly linked to needing intermediation: a sizeable contingent of farmers visiting wakalas frequently do not need help and are just there to get airtime refilled, some farmers who visit would be the target of intermediation, and some farmers would never visit the wakala due to their low use or understanding of technology. Our HH survey shows broad heterogeneity in agricultural workers’ comfort with USSD: those who need help are evenly male and female and of varying ages. Pilot results mirrored this heterogeneity; while increased comfort was not correlated with more complex use, pHHs used the app more and more complexly than bHHs. Therefore, with well designed intermediation schemes involving wakalas, non-users due to lack of trust and lack of technology literacy may be able to access information while growing their comfort in USSD. 


\section{Conclusion}

We describe the design and scaling of a dual-platform technology for agricultural directory dissemination in Sub-Saharan Africa. Through a household survey, we find that smartphone penetration remains low in periurban and rural Tanzania, necessitating the continued development of usable and trust-conferring feature phone technologies to support a majority of farmers. We show that USSD apps are preferred by both feature phone and smartphone owners, although scaling a directory that relies on USSD has limited marginal gains for users due to bounds on the underlying USSD protocol itself. In contrast, Android apps allow for more complex use and easier information gain. Our small pilot study in intermediation of the USSD application using wakalas shows promise for bridging the gap to users that report low comfort with technology and therefore are left out of the expanded information ecosystem.


\begin{acks} 

We give heartfelt thanks to our collaborators at Cornell University for giving us the opportunity to be involved in extended fieldwork and our partners at EDI Global who enumerated the HH survey, facilitated user testing sessions, etc. The work would not have been possible without their expertise. 

\end{acks}


\nocite{*} 
\bibliographystyle{ACM-Reference-Format.bst}
\bibliography{references} 

\appendix
\section{Appendix: Training}

\balance
Here is a summary of the English version of the script  used for enrolling participants in the study. We present the script to both document the enrollment processes and to provide additional background on the intervention. This script was used for household members invited to attend a public meeting. The script was used by the Tanzanian researchers conducting the enrollment sessions. The Swahili version of the script was used for enrollment.

The process began with an introduction of the purpose of the study: ``Thank you for answering my questions, we are very grateful for your assistance with this study. I would like to tell you more about this project. We are interested in learning whether some additional information about businesses in Kagera would be useful for farming households. To better understand this, we are giving some of our participating households access to a phone-based telephone directory. There is no specific reason that your household was chosen to receive this directory – my computer randomly tells me at the end of the interview whether a respondent is eligible to access the directory service.''

The navigation system of the menus was introduced with the scenario of a farmer wanting to find a transporter for farm produce. This motivated the service and also identified various options such as village delivery or paying with mobile money. This was done step by step, beginning with:
\begin{itemize}
\item ``First, dial *149*26\#, and show them the initial screen. Explain that there are 2 ways to search. The first way by navigating menus (you can navigate the menus by selecting a sector (option \#1), or by selecting a location (option \#2)). The second way is by typing what you want to find (option \#3) There is also a help button, option 4.''
\item ``To find a transporter, select \#1: Tafuta kwa kuchagua. Say that there are six main sectors in the directory. [show these on the USSD menu as you describe them, mention that you can select \#0: for the `next' page]: wholesale traders; retailers, including shops and kiosks selling all kinds of goods; transporters, including boda bodas, lorries, and others; agricultural processors, such as those with milling machines; skilled tradespeople (fundi), including tailors, carpenters, mechanics, and others; and businesses that provide services, such as restaurants, salons, wakalas, and financial institutions. Those are just some examples, there are many other types of businesses in those sectors.''
\item  ``When on the second page of sectors, mention that they can press 99 to return back at any time.
Press 99 to return to the 1st page of sectors.''
\end{itemize}
The walkthrough then continued to describe the geographic hierarchy and the placement of business in different organizational units, introducing the key organizational principle for the directory.

The script then described how the directory was constructed by the research team, emphasizing how this work was done with the local VEOs (Village Executive Officers).  All of the villages were visited by the research team, and there was an attempt to build confidence in the relevance of the information.

The script continued to encourage different types of usage on the directory: ``There are many potential uses of this directory. Let me give you some examples. First, suppose you are interested in learning whether a certain kind of seed is available for purchase. Instead of traveling to a business to find out, you can look up the phone numbers of businesses that sell agricultural inputs, and call them first. You could call a few different businesses in nearby towns, to see if they have the seed in stock, to ask the price, and to see if they stock other products that you need.''  and then continued with another specific example of finding a seed shop by searching on business type.  A third example was used around selling crops, with the motivation of finding the best price through calling different types of businesses such as wholesale traders and retail shops.

The script then had the participants explore that application and ask questions.   The first discussion was strictly about the USSD application, and at this point the smart phone application is introduced, by saying that you can put an application on a smartphone that holds all of the information.  The first discussion of the smartphone application is around data connectivity with the point that once it is loaded on the phone, it does not need connectivity to be used.

The final part of the script returned to the logistics of the USSD application:  ``The USSD service and the mobile app are restricted – they are only available to participants in this study. It will take us about a week to add your phone number to the system, so that you can access these services. We will send you a text message when the services are available to you.'' and then talked about the actual study and the need to collect usage data and the confidentiality policy,  It was then emphasized that the USSD shortcode is completely free, and the application will not consume airtime credit.
An information sheet was then given to participants.


\end{document}



Here is a summary of the English version of the script  used for enrolling participants in the study. We present the script to both document the enrollment processes and to provide additional background on the intervention. This script was used for household members invited to attend a public meeting. The script was used by the Tanzanian researchers conducting the enrollment sessions. The Swahili version of the script was used for enrollment.

The process began with an introduction of the purpose of the study:  ``Thank you for answering my questions, we are very grateful for your assistance with this study. I would like to tell you more about this project. We are interested in learning whether some additional information about businesses in Kagera would be useful for farming households. To better understand this, we are giving some of our participating households access to a phone-based telephone directory. There is no specific reason that your household was chosen to receive this directory – my computer randomly tells me at the end of the interview whether a respondent is eligible to access the directory service.''

The directory was then introduced with the scenario of locating a transporter for agricultural goods.
Let me show you how to use the directory through navigating the menus. Suppose that you would like to find a transporter for your goods. If you cannot get what you need from a nearby area, you can call transporters in the directory to see if anyone can meet your needs. That way you can use the directory to find what you need, instead of searching in multiple locations. You may even decide order goods for delivery to the village, and to pay with mobile money, rather than traveling to buy it. A transporter can help you with this.

• First, dial *149*26\#, and show them the initial screen. Explain that there are 2 ways to search. The first way by navigating menus (you can navigate the menus by selecting a sector (option \#1), or by selecting a location (option \#2)). The second way is by typing what you want to find (option \#3) There is also a help button, option 4.
Here is how to find a transporter:
• Select \#1: Tafuta kwa kuchagua. Say that there are six main sectors in the directory. [show these on the USSD menu as you describe them, mention that you can select \#0: for the “next” page]: wholesale traders; retailers, including shops and kiosks selling all kinds of goods; transporters, including boda bodas, lorries, and others; agricultural processors, such as those with milling machines; skilled tradespeople (fundi), including tailors, carpenters, mechanics and others; and businesses that provide services, such as restaurants, salons, wakalas, and financial institutions. Those are just some examples, there are many other types of businesses in those sectors.
• When on the second page of sectors, mention that they can press 99 to return back at any time.
• Press 99 to return to the 1st page of sectors
• Select \#3. Usafirishaji
• Select subsector: Boda/pikipiki
• Select the district you are in
• Select any village available on the page
• Explain what “Biashara zote” means - all businesses that are Boda/pikipiki in this village. They can select biashara zote, or they can select a subvillage to further narrow down the search.
• Select \#1: Biashara zote
• Select the first business in the list
• When you select a specific business, the information provided is: (1) business name, or the name of the district/village/sector; (2) subvillage; (3) a subsector of the business, which might highlight areas of speciality, and (4) the main phone number for the business.

The information in this directory was assembled by our research team, through a survey that we conducted a few months ago. We worked with the support of local VEOs, and every business listed here was visited by one of our team members. It is possible that some businesses will close this year, or change their business practices. But we hope and expect that the directory will remain very relevant for years to come.

There are many potential uses of this directory. Let me give you some examples. First, suppose you are interested in learning whether a certain kind of seed is available for purchase. Instead of traveling to a business to find out, you can look up the phone numbers of businesses that sell agricultural inputs, and call them first. You could call a few different businesses in nearby towns, to see if they have the seed in stock, to ask the price, and to see if they stock other products that you need.

Show Example:
• Begin a new session by dialing *149*26\#
• Let them know that you will show them how to find an Agrovet, a shop that sells seeds, by search by typing
• Select \#3 - Tafuta kwa kuandika
• Explain the 4 things that they can type in: Business Name, Location, Product/Crop, Sector/Service, or Owner Name
• Select \#3 Product/Service
• Select Subproduct = Bidhaa za Kilimo
• Type in mbegu
• Select \#2: Hapana to sorting by location
• Select the first business screen
You can select 0 to see the next business in the search results. (demonstrate this)

For a third example, think of the next time that you will have crops for sale. To make sure that you get a good price, you can call wholesale traders, retail shops, and anyone else in the directory that you think might have information about prices. This could help you choose a new buyer or improve your bargaining position with local buyers.

• First, dial *149*26\#
• Select \#1: Tafuta kwa kuchagua.
• Select \#4. Wafanyabiashara wa jumla
• Select the district you are in
• Select any village available on the page
• Explain what “Biashara zote” means - all businesses that are Wafanyabiashara wa jumla in this village
• Select \#1: Biashara zote
• Select the first business in the list
• Describe how the business phone number is listed, along with the owner name, subsector, subvillage, etc.

Those are just three examples. There are of course many other uses of this directory, for arranging transport, finding a wakala agent, making plans with a salon, finding a new tailor, and others. Let's look at businesses in a village near here. [Ask the name of a nearby trading center, and look it up together. In larger villages you can check whether the village itself is represented in the directory.] Feel free to take some time, look through the service on my phone, and ask me any questions. [Let the respondent browse through the USSD directory]. Do you have any questions for me? [Answer questions as needed].
If you or someone in your household has a smartphone, or plans to get a smartphone in the next year, there is a second way that you can access the directory. We have developed an app that contains the same information as the USSD directory. [Open the app on your phone]. Once you download the app, it works without an internet connection. It will only use a small amount of data, about once a week, to exchange information with the server and check for updates. Let me show you how it works. [Open the app and navigate to some of the same firms used in the USSD example]. Feel free to try it out. Do you have any questions? If there is a strong internet connection here, would you like me to show you how to find the smartphone app at the Google Playstore?

The USSD service and the mobile app are restricted – they are only available to participants in this study. It will take us about a week to add your phone number to the system, so that you can access these services. We will send you a text message when the services are available to you.  
Because we are interested in the information needs of farming households, we will collect data from both the USSD service and the mobile client. We want to see what people search for, what firms they visit, and other ways that they use these services. As with the survey data, any information we collect about how households use these services will be kept strictly confidential, and will never be shared in a way that would allow someone to learn your phone number or identity.  
I want to conclude with just a few important points and reminders. 

1. The USSD shortcode is completely free. It will not consume any of your data or airtime credit. The smartphone app is also free to use, and uses only a tiny bit of data after you download it.
 
2. We will text you when you have access to these services. However, sometimes our text messages do not arrive because of network problems. If you don't get a text from us, try to access the directory every few days.
 
3. We are trying to maintain a quality directory. But some business phone numbers may change or become unavailable. So, don't give up if you try a couple of numbers and they are not available or not correct. These are all Kagera-based businesses that signed up earlier this year.
 
I want to encourage you to use these services in whatever way is best for you and your household. We plan to return some time next year. When we do, we’ll ask you some more questions about the activities of this household. Until then, thank you again for taking time to speak with me. This information sheet has more details about the project. 
[Thank the respondent and end the interview]